# Virtues of Including Hydrogen in the Patterns of Element Abundances in Solar Energetic Particles

**Donald V. Reames**

Institute for Physical Science and Technology, University of Maryland, College Park, MD 20742-2431 USA, email: dvreames@umd.edu

**Abstract** We revisit a multi-spacecraft study of the element abundances of solar energetic particles (SEPs) in the 23 January 2012 event, where the power-law pattern of enhancements *versus* the mass-to-charge ratio $A/Q$ for the elements C through Fe was partly disrupted by a break near Mg, which turned out to be an unfortunate distraction. In the current article we find that extending that least-squares fits for C – Fe down to H at $A/Q = 1$ lends much more credence to the power laws, even though H itself was not included in the fits. We also investigate the extent of an adiabatically invariant "reservoir" of magnetically-trapped particles behind the shock wave in this event.

# 1. Introduction

Many years ago it was shown that the abundances of the elements with atomic numbers 6 ≤ Z ≤ 30 in large solar energetic particle (SEP) events, relative to a corresponding reference (coronal) abundances for these events, could increase or decrease as a power law in the mass-to-charge ratio *A/Q* of the ions (Breneman and Stone, 1985). Since the pattern of *Q*-values for the elements depends upon the electron temperature of the ion source plasma, this has become a means of actually estimating this source temperature by selecting the best-fit power law (Reames, 2015, 2016, 2017a, 2018b). While this technique seems to be broadly applicable to large "gradual" SEP events, a multi-spacecraft study comparing power-law abundance patterns for events, observed simultaneously by *Wind* and the two *Solar TErrestrial RElations Observatory* (STEREO) spacecraft, was stymied by the event of 23 January 2012 where an apparent break in the power law near Mg, seen at all three spacecraft (Reames, 2017b), led to questions of the applicability of the technique to this event.

More-recent studies have extended *A/Q* techniques to include H (Reames, 2019), a significant extension in *A/Q* from 2 down to 1. For most gradual events, especially the larger ones, H usually agrees well with the extension of the power-law fit line of abundance enhancements *versus A/Q* from the elements with Z ≥ 6 (Reames, 2019, 2020).

There are two primary sources of SEP events, historically designated "impulsive" and "gradual" (*e.g.* Reames, 1988, 1995b, 1999, 2013, 2017a; Gosling, 1993). In small "impulsive," or $^3$He-rich, SEP events (Mason 2007; Bučík 2020), acceleration occurs at sites of magnetic reconnection in solar jets (Kahler, Reames, and Sheeley, 2001; Bučík *et al.*, 2018a, 2018b; Bučík 2020). Here, element abundances, relative to those in the corona, are enhanced as a steeply increasing power of *A/Q*, increasing a factor of ≈1000 across the periodic table from H or He to Pb (Reames, 2000; Mason *et al.*, 2004; Reames and Ng, 2004; Reames, Cliver, and Kahler, 2014a, 2014b) during the reconnection itself (*e.g.* Drake *et al.*, 2009). On average, abundance enhancements increase as $(A/Q)^{3.64 \pm 0.15}$ with *A/Q* determined at a temperature $T \approx 3$ MK in impulsive SEP events (Reames, Cliver, and Kahler, 2014a).

By way of contrast, the larger "gradual" SEP events (Lee, Mewaldt, and Giacalone, 2012; Desai and Giacalone, 2016), involve acceleration at shock waves, driven by fast, wide coronal mass ejections (CMEs; Kahler *et al.*, 1984; Lee, 1983, 2005; Reames, Barbier, and Ng, 1996; Zank, Rice, and Wu, 2000; Cliver, Kahler, and Reames, 2004; Ng and Reames, 2008; Gopalswamy *et al.*, 2012; Cohen *et al.*, 2014; Kouloumvakos *et al.*, 2019), that usually sample ambient coronal material (Reames, 2020). Coronal abundances sampled by SEPs differ from those in the photosphere by a factor that depends upon the first ionization potential (FIP) of the element. High-FIP (>10 eV) elements travel as neutral atoms across the chromosphere while low-FIP elements are ionized initially and are preferentially boosted upward by a factor of ≈3 in intensity by the action of Alfvén waves (Laming, 2015; Reames, 2018a; Laming *et al.*, 2019); these FIP-processed coronal ions form the reference abundances sampled as SEPs much later (Webber, 1975; Meyer, 1985; Reames, 1995a, 2014). After acceleration, ion scattering during transport can depend upon a power of the ion magnetic rigidity, and hence abundances depend upon $A/Q$ for ions compared at a constant velocity (Parker, 1963; Ng, Reames, and Tylka, 1999, 2001, 2003, 2012; Reames, 2016, 2019), leading to the observed power-law dependence. For example, since Fe scatters less than O, Fe/O is often enhanced early in an event and therefore depleted later, producing a power-law dependence upon $A/Q$ that increase early and decrease later (*e.g.* Breneman and Stone, 1985, Reames, 2016, 2019).

The distinction of impulsive and gradual SEP events becomes complicated when local shock waves at CMEs in jets reaccelerate the impulsive ions pre-accelerated in the magnetic reconnection, and also, large pools of impulsive suprathermal ions can collect near active regions with many jets, producing [3]He-rich, Fe-rich background levels often seen (Desai *et al.*, 2003; Bučík *et al.*, 2014, 2015; Chen *et al.*, 2015). These pools of impulsive suprathermal seed ions are preferentially accelerated (Desai *et al.*, 2003; Tylka *et al.*, 2005; Tylka and Lee, 2006) and even dominate SEPs with weaker shocks that need this boosted injection, while stronger shocks are more-easily seeded by ambient coronal plasma. Reames (2020) has identified four paths for element abundances: (i) the "pure" impulsive events, SEP1, (ii) impulsive events with shocks, SEP2, (iii) gradual events dominated by impulsive seed ions, SEP3, and (iv) strong gradual events dominated by ambient corona, SEP4. We might expect that events typically seen by multiple, widely-separated spacecraft would tend to be the strong SEP4 events where the H generally fits on the abundance power law with the other elements (Reames, 2020).

Another especially interesting phenomenon is observed behind the shock wave in large gradual events: the SEP "reservoir." Reservoirs contain magnetically-trapped, invariant populations of particles of interest because particle intensities decrease adiabatically, maintaining their spectral shape and abundances, as the magnetic bottle containing them expands (see Sect. 5.7 of Reames 2017a; Reames, Kahler, and Ng, 1997). Early measurements on the *Pioneer* spacecraft found comparable intensities of ≈20 MeV protons extending ≈180° in longitude around the Sun late in SEP events (*e.g.* McKibben, 1972). Later, equal intensities were found in large events over 2.5 AU radially between *Ulysses* and IMP 8 near Earth by Roelof *et al.* (1992) who coined the term "reservoir."

*A priori* one might expect higher-rigidity or higher-velocity particles to scatter less, to sample boundaries more often, penetrate them more easily, and to slowly leak away, so that spectra steepen and abundance ratios, like Fe/O, decrease with time. However, this does not happen within reservoirs, where spectra and abundances tend to be invariant in space and time, as if all intensities decrease in unison as the volume of the magnetic bottle increases with time. Particle transport is nearly scatter-free within reservoirs as we see when newly-injected ³He-rich events traverse them (Mason *et al.*, 1989; Reames, 1999, 2013, 2017a). It seems especially amazing to the author that SEPs are so well contained at CMEs and reservoirs, while CMEs and even magnetic clouds are so easily penetrated and filled from outside by anomalous cosmic rays (ACRs; Reames, Kahler, and Tylka, 2009) at solar minimum, although these latter CMEs do lack shocks. Are these regions magnetically open or closed? This makes reservoirs sufficiently interesting to take this opportunity to extend our study of the reservoir in the 23 January 2012 event and compare it with reservoirs seen in previous solar cycles (*e.g.* Reames, Kahler, and Ng, 1997; Lario, 2010; Reames, Ng, and Tylka, 2012).

The original study of the power-law abundance pattern of the 23 January 2012 event that we would like to revisit is that of Reames (2017b), comparing the widely separated platforms: STEREO-Ahead (A) and -Behind (B) Earth in solar orbit, and the *Wind* spacecraft near Earth. During this large SEP event, we now include the abundance of H with the relative abundances of the elements He, C, N, O, Ne, Mg, Si, S, Ar, Ca, and Fe from the *Low-Energy Matrix Telescope* (LEMT; von Rosenvinge *et al.*, 1995) on *Wind* and the *Low-Energy Telescopes* (LETs; Mewaldt *et al.*, 2008; see also Luhmann *et al.*, 2008) on STEREO. Abundances are taken primarily from the 3.2–5 MeV amu$^{-1}$ interval on LEMT, although H is only available near 2.5 MeV amu$^{-1}$. We use the 4–6 MeV amu$^{-1}$ interval for the ions on LET (S is not available on LET; all LET data are from http://www.srl.caltech.edu/STEREO/Level1/LET_public.html). Abundance enhancements are measured relative to the average SEP abundances listed by Reames (2017a; see also Reames, 1995, 2014, 2020).

## 2. The 23 January 2012 SEP Event

The source temperature analysis for *Wind* observations of this event is shown in Figure 3 of Reames (2018b), while Figure 4 of Reames (2017b) shows the comparative analysis of STEREO A (without H). A similar analysis of STEREO B data (Figure 5 of Reames, 2017b) breaks down because of the relatively flat pattern of abundances *versus A/Q* early and especially because of the poor statistics of the STEREO B data later in the event. Source temperatures can only be determined when the relative abundances are either rising or falling *versus A/Q*. This time, we surmount the problem of flat abundances *versus A/Q* simply by assuming for STEREO B the well-determined $T \approx 1.6$ MK seen on both *Wind* and STEREO A (Reames, 2017b).

Panels (c), (d), and (e) of Figure 1 compare the time evolution of the power-law fits of the relative abundance enhancements *versus A/Q* for the three spacecraft. *Wind* data are spaced at 8-hr intervals while the STEREO data are for daily intervals, with the beginning time of each interval listed to the right of the fitted data.

Judging the fitted data in Figure 1, it is true that some time periods seem to show a systematic departure from the power-law fit, if we view the $6 \leq Z \leq 26$ data alone; note, for example, the bottom fits in Figure 1(d). However, it is also true that the power-law fits based upon the $6 \leq Z \leq 26$ elements project remarkably close to H at $A/Q = 1$. H is not included in any of these fits. Thus, we must say that despite some modest high-$Z$ departures, the power-law behavior dominates the abundances. If we know the abundances of heavy elements C – Fe at any time and place in this event, we can fairly reliably predict the abundance of H at the same velocity.

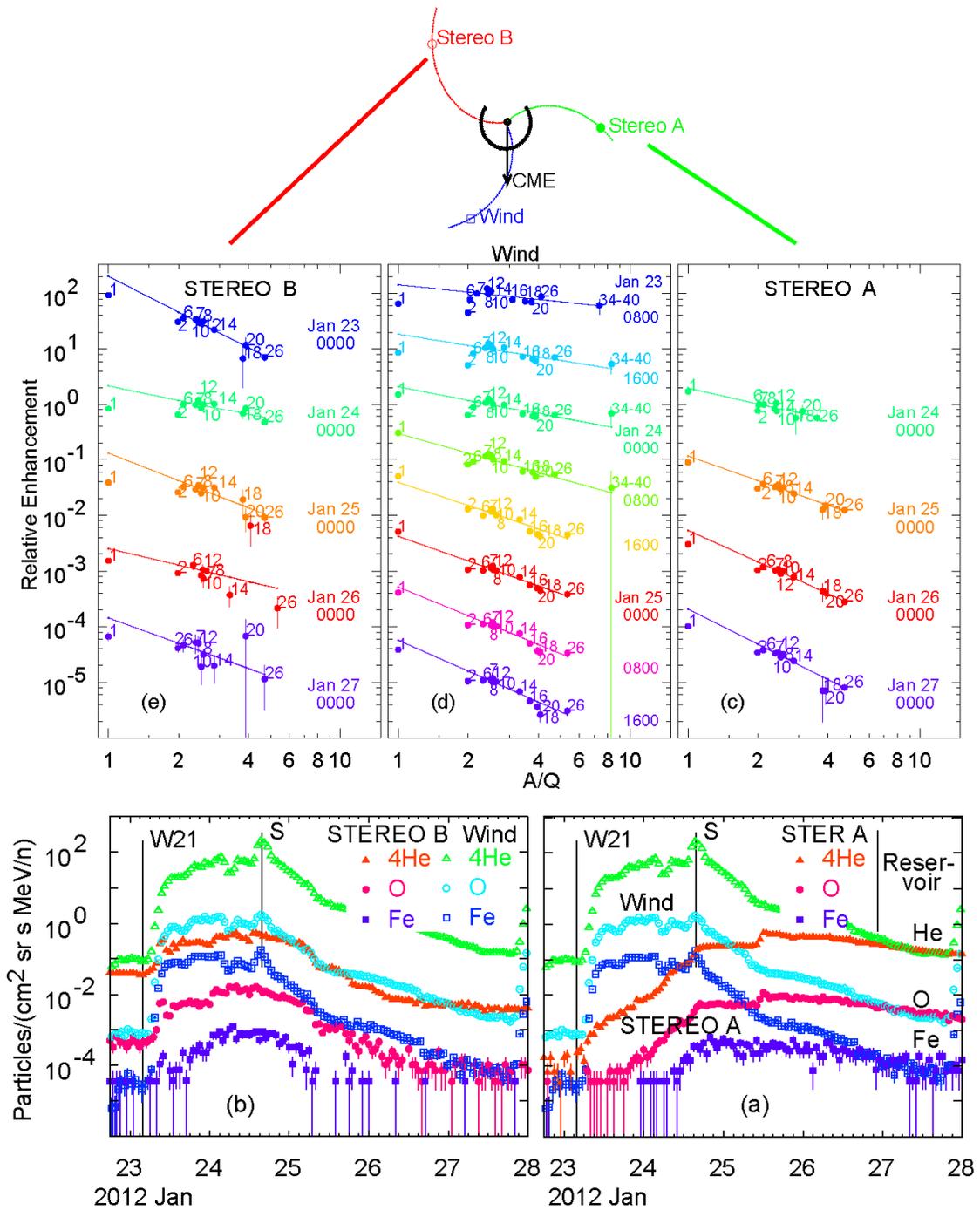

**Figure 1** Intensities (*lower panels*) of $^4$He, O, and Fe on *Wind* are compared with those on (**a**) STEREO A and (**b**) STEREO B. *Upper panels* show relative enhancements, labeled by *Z*, versus *A/Q* with the best-fit power law for elements with $Z \geq 6$ extrapolated down to H at $A/Q = 1$, displaced at each listed time for (**c**) STEREO A, (**d**) *Wind*, and (**e**) STEREO B. The sketch at the *top* of the figure shows the spacecraft locations around the Sun relative to the downwardly directed CME-driven shock.

Figure 1(a) shows that a reservoir slowly evolves to include both *Wind* and STEREO A late on 26 January where the intensities and spectra of all element species become equal. At this time the shock wave would have moved well beyond the spacecraft to about 2 AU. Actually, the slopes of the power-law fits for *Wind* and STEREO A begin to stabilize and agree as early as 25 January. The reservoir does not extend around as far as STEREO B.

Does this reservoir only exist so late in the events when it is seen by both spacecraft? It is also possible to detect a reservoir on a single spacecraft. If we normalize intensities of particles of different species and different energies at a single time in a reservoir, those intensities will stay normalized at later times, because the spectral shapes and abundances are invariant.

Figure 2 shows intensities for a variety of particle species and energies available on the *Wind* spacecraft, all normalized at 1000 UT on 25 January 2012. The particle intensities show significant invariant behavior for much of the time after the shock passage, until they reach background or lack adequate statistics. The reservoir existence is not limited to the tiny time interval on 27 January where *Wind* and STEREO A overlap, as seen in Figure 1(a).

**Figure 2** Intensities, at *Wind*, of the particles and energies listed, are normalized at 1000 UT on 23 January 2012 and plotted *versus* time.

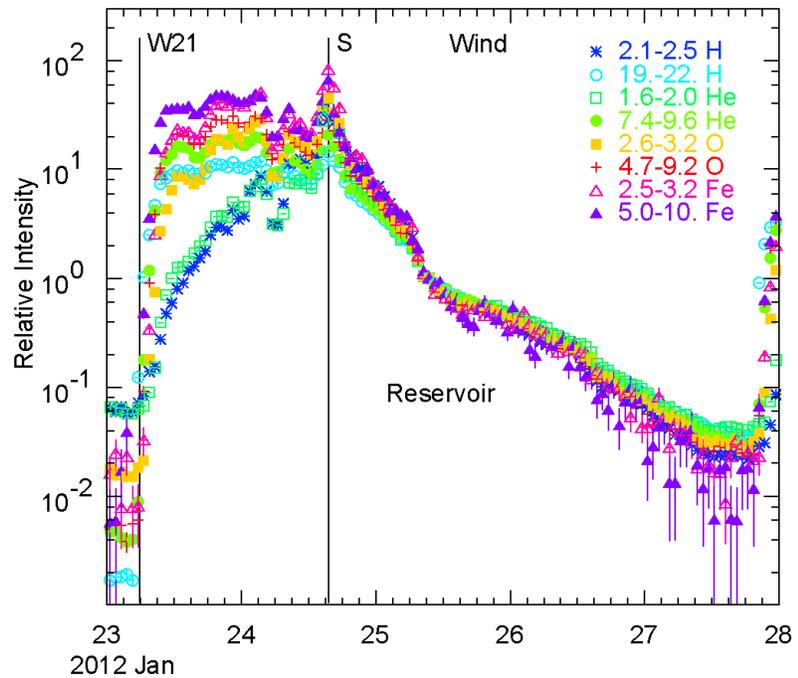

# 3. Discussion and Conclusions

Adding H abundances can significantly increase the leverage for determining the power-law behavior for those SEP4 events (Reames, 2020) where the H abundance is usually part of the power law. For the SEP3 events, with "excess protons" from ambient plasma in addition to enhanced $Z > 2$ ions reaccelerated from impulsive suprathermal ions, H abundances would not help because H does not fall on the power law. However, the weaker SEP3 events are less likely to be studied at all on widely separated spacecraft. For example the event of 4 August 2011 (Cohen *et al.*, 2014) can be well studied on *Wind*, while on STEREO A the event is still clearly Fe-rich but it is too small to define power-law abundances. Using different techniques, Cohen *et al.* (2014) have concluded that longitudinal spatial transport is not rigidity dependent, an additional confirmation of this finding that was first proposed for spatial distributions by Mason, Gloeckler, and Hovestadt (1984) to justify shock acceleration. As seen in Figure 1, rigidity dependence, *i.e.* slope of the power law, does tend to steepen with time at a given longitude early in the event as higher-rigidity ions flow away because of their reduced scattering.

Individual SEP events can be quite complex. In general, we have invoked this simplifying assumption that the behavior of SEP ions will depend upon their velocity and as a power law upon their rigidity. This assumption seems to work for a surprising number of cases and for broad variety of categories of SEP events, although exceptions do exist (Reames, 2020). For example, streaming protons can modify Alfvén waves that subsequently affect ion scattering and disrupt simple power laws in rigidity (Ng, Reames, and Tylka, 1999, 2003; Reames, Ng, and Tylka, 2000; Reames and Ng, 2010). We certainly do encounter events where the simple power-law assumption does not work (*e.g.* Reames 2020).

However, when searching for power laws in *A/Q*, the increased leverage gained by the additional factor of two from including protons, can be critical. In our previous study (Reames, 2017b) the omission of protons, typical before 2019, led to an unfortunate focus on less significant variations. Proton abundances have greatly improved this picture for the 23 January 2012 SEP event as seen in Figure 1. More generally, proton abundances that do or do not agree with fits from those of heavier ions have provided a pow-erful new measure of SEP physics of shock acceleration and its seed population (Reames, 2019, 2020).

Considering reservoirs, the regions of multi-spacecraft overlap of reservoirs in the STEREO era (*e.g.* Cohen 2014; Reames 2017b) seem to be much smaller than those we saw in the *Helios* – IMP 8 – *Voyager* era (Reames, Kahler, and Ng, 1997; Reames, Ng, and Tylka, 2012).   However, this is partly because the *Helios s*pacecraft were typically separated by 30° – 45° in longitude with several spacecraft within <90° of the CME source longitude.  The STEREO spacecraft were launched just prior to a long solar minimum, and they separated rapidly, so that significant SEP events were not seen until the spacecraft were each ≈120° from Earth.  Thus we suspect that reservoirs are just as potent in Solar Cycle 24 as they were in Solar Cycle 21, and we have seen that they do extend over 120° from their source, relatively late in an event.  We are not yet sure why CME-based magnetic structures would contain SEPs so well, but exclude ACRs so poorly.  However, it is possible that turbulence at the shocks that accelerate SEPs, but are missing for ACRs, play a greater role than magnetic structure of CMEs, in preventing outward SEP escape.  Converging fields reflect particles near the Sun.

## Disclosure of Potential Conflicts of Interest

The author declares he has no conflicts of interest.